\documentclass[aps,pra,twocolumn,showpacs]{revtex4}

\usepackage{amsmath}
\usepackage{graphicx}
\usepackage{epsfig}

\allowdisplaybreaks

\newcommand{\Av}[1]{\left\langle #1 \right\rangle}

\newcommand{\st}[1]{\left| #1 \right\rangle}
\newcommand{\stdag}[1]{\left\langle #1 \right|}

\def\mathand{ \quad \mbox{and} \quad }

\begin{document}

\title{Many particle entanglement in two-component Bose-Einstein Condensates}

\author{A.~Micheli$^{1}$, D.~Jaksch$^{1}$, J.~I.~Cirac$^{1,2}$, and P.~Zoller$^{1}$}

\affiliation{ $^1$Institut f\"ur Theoretische Physik, Universit\"at Innsbruck,
A-6020 Innsbruck, Austria\\
$^2$ Max--Planck Institut f\"ur Quantenoptik, Hans--Kopfermann Str. 1, D-85748 Garching, Germany }

\date{\today}

\begin{abstract}
We investigate schemes to dynamically create many particle
entangled states of a two component Bose-Einstein condensate in a
very short time proportional to $1/N$ where $N$ is the number of
condensate particles. For small $N$ we compare exact numerical
calculations with analytical semiclassical estimates and find very
good agreement for $N \geq 50$. We also estimate the effect of
decoherence on our scheme, study possible scenarios for measuring
the entangled states, and investigate experimental imperfections.
\end{abstract}


\maketitle

\section{Introduction}
\label{Introduction}

The creation of many particle entangled states in macroscopic
systems is one of the major goals in the studies on fundamental
aspects of quantum theory
\cite{Monroe4,Haroche1,Bouwmeester,Polzik1,Bennett1Measures}. The
notion of entanglement in macroscopic ensembles allows to
investigate the boundary between quantum physics and classical
physics and, possibly, could also give some insight into the
measurement process \cite{Einstein1,GHZreview,BellIneq1}. The
experimental creation of many particle entanglement could also lead
to the realization of several of the ``Gedankenexperiments''
proposed in the early days of quantum theory
\cite{ErwinSchrodinger}. Also, apart from the fundamental physical
interest in entanglement, the whole field of quantum computing and
quantum information is based upon the ability to create and control
entangled states \cite{PhysWorld}.

The experimental achievement of atomic Bose-Einstein condensation
(BEC) \cite{BEC} has opened fascinating possibilities for studying
quantum properties of a macroscopic number of cold quantum
degenerate atoms in the laboratory. Interesting aspects of many
particle entanglement can be studied by using condensates with
internal degrees of freedom \cite{Sorensen01,Pu,
Duan,Duan2,Bigelow,Poulsen1,Helmerson1}. For instance, it has been
shown that the coherent collisional interactions in a BEC allow to
generate substantial many-particle entanglement in the spin degrees
of freedom of a two-component condensate \cite{Sorensen01} during
the free evolution of the condensates.

The extremely long coherence time in a BEC is one of the key
features making the proposals for engineering many particle
entanglement in a BEC feasible \cite{CornellDeco}. However, one
still has to make sure that (i) the creation of the many particle
entangled state takes place on a time scale much shorter than the
coherence time, and (ii) the produced many particle entangled
states are robust against decoherence, i.e., particle loss should
not destroy the entanglement.

In this paper we will investigate in detail a scheme that allows
to create many particle entangled states in a two-component BEC
interacting with a classical laser field or microwave field on a
time scale proportional to $1/N$ with $N$ the number of condensate
particles (see also \cite{GordonSavageCat}). Let us briefly
explain the basic idea of this scheme. If we concentrate on the
dynamics of the internal states of the condensate (e.g. atomic
hyperfine levels) the Hamiltonian $H$ of the system is given by
(cf.~Sec.~\ref{TMHam})
\begin{equation}
H=\chi S_z^2+\Omega S_x, \label{Hamiltonian}
\end{equation}
where $\chi$ is determined by the interaction strengths between
the condensate particles and $\Omega$ is the Rabi frequency of the
external field interacting with the condensate. Here the angular
momentum operator ${\bf S}=\{S_x,S_y,S_z\}$ is defined by
\begin{eqnarray}
S_x &=& \frac 1 2 (a^\dagger b + b^\dagger a), \nonumber \\
S_y &=& \frac i 2 (b^\dagger a - a^\dagger b), \nonumber \\
S_z &=& \frac 1 2 (a^\dagger a - b^\dagger b), \nonumber
\label{AngularMomentumTwoMode}
\end{eqnarray}
where $a$ ($b$) are bosonic destruction operators for particles in
internal state $A$ ($B$) with a fixed spatial mode function
$\psi_{A(B)}$. We consider the situation, where initially all the
condensate particles are in internal state $\st{A}$. This state
corresponds to an eigenstate of $S_z$ with eigenvalue $N/2$. Then a
$\pi/2$ pulse is applied to the condensate which brings each
particle in a superposition state $(\st{A}+\st{B})/\sqrt 2$
corresponding to an eigenstate of $S_x$ with eigenvalue $N/2$. The
internal wave function $|\Psi\rangle$ of the condensate is thus
given by
\begin{equation}
\st{\Psi(t=0)} = \frac{1}{\sqrt{2^N N!}}
\left(a^\dagger+b^\dagger\right)^N \st{vac}, \label{StartingPsi}
\end{equation}
where $\st{\rm{vac}}$ is the vacuum state.

\begin{figure}
\begin{center}
\epsffile{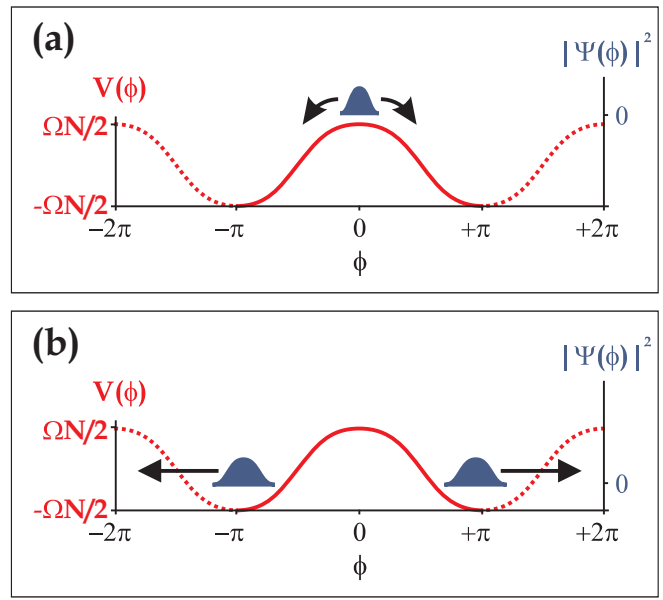}
\end{center}
\caption{Schematic time evolution using the phase-model. (a) The
initial state is a Gaussian wave packet centered at a maximum of
the phase potential. (b) The time evolution according to $H_\phi$
splits the wave packet into a superposition of two wave packets
moving in opposite direction. \label{fig:Split_pic}}
\end{figure}

To get a qualitative understanding of the subsequent time
evolution of $\st{\Psi(t)}$ we use the familiar phase model
\cite{Menotti1,Jaksch1} where one replaces $S_z \rightarrow -i
\partial_\phi$ and $S_x \rightarrow N/2 \cos \hat \phi$, with $\phi$
the relative phase between the two condensate components, in the
Hamiltonian and finds (cf.~Sec.~\ref{PhaseModel1})
\begin{equation}
H_\phi = -\chi \frac{\partial^2}{\partial \phi^2} + \frac{\Omega N}
2 \cos \hat \phi. \label{HamiltonPhase}
\end{equation}
We note that the eigenstates of $\hat \phi$ given by $|\phi\rangle
\sim \sum_m e^{-i m \phi}|N/2-m\rangle_A |N/2+m\rangle_B$ are
entangled states of particles in  states $A$ and $B$
\cite{Zwerger1, Leggett1, Jaksch1}. Projection on these eigenstates
gives the phase wave function $\Psi(t,\phi)=\langle \phi
\st{\Psi(t)}$. For large condensate particle number $N$ the phase
wave function corresponding to the initial state $\Psi(0,\phi)$ is
well approximated by a narrow Gaussian wave-packet with a width of
$\sigma=1/\sqrt{N} $ centered at $\phi=0$, i.e.~a maximum of the
phase potential $V(\hat\phi)=\Omega S \cos \hat\phi$ as
schematically shown in Fig.~\ref{fig:Split_pic}a. The time
evolution due to $H_\phi$ will first squeeze this wave packet due
to the harmonic terms of $V(\hat\phi)$ resulting in an increased
width. Then the anharmonicity of $V(\hat\phi)$ becomes important
and the wave packet splits up as shown in
Fig.~\ref{fig:Split_pic}b. These two wave packets correspond to
wave packets of opposite relative particle number (measured by
$S_z$) since they move in opposite direction. Their superposition
is thus a macroscopically entangled state. In the remainder of this
paper we will investigate the process described above
quantitatively. In particular, we will use a semiclassical
approximation which allows to analytically estimate properties of
the many particle entangled states like the positions of maxima in
the relative particle number distribution.

The paper is organized as follows: In Sec.~\ref{Model} we will
introduce the model. We derive a two mode Hamiltonian describing a
two component condensate interacting with a laser or microwave
field. Then we discuss the phase model already used in the
introduction and present a semiclassical model. In
Sec.~\ref{Overview} we define the most common types of many
particle entangled states and give a short overview of recent
proposals for entanglement creation in BEC's. We present our
results in Sec.~\ref{sec:SCRes} and Sec.~\ref{sec:numres} where we
compare the properties of the many particle entangled states found
by numerical solution of the exact Schr\"odinger equation with the
semiclassical estimates. Furthermore in
Sec.~\ref{sec:One-Axis-Pre-Squeezing} we improve the scheme by
one-axis-pre-squeezing. Finally, in Sec.~\ref{Disc} we discuss the
stability of the many particle entangled states under decoherence
and investigate possible measurement strategies to identify the
entangled states. Also we study the influence of imperfections in
the external field on our scheme.

\section{Model}
\label{Model}

In this section we present the model used to study the two
component condensate interacting with an external field. We write
down the Hamiltonian for this system and use the two mode
approximation to obtain a simplified description of the system in
terms of angular momentum operators. Then we introduce the phase
model and a semiclassical model.

\subsection{Hamiltonian}
\label{TwoModeSection}

We consider a two component BEC consisting of $N$ atoms in two
different hyperfine states $\st{A}$ and $\st{B}$ coupled by a
Raman laser or microwave field (internal Josephson effect
\cite{Zwerger1, Leggett1, Jaksch1}). The Hamiltonian of this
system is given by
\begin{equation}
H = H_A+H_B+H_{\rm{ext}},
\end{equation}
where $H_A$ and $H_B$ describe the two component condensate and
$H_{\rm{ext}}$ the interaction with the external field. In second
quantization the terms are given by ($\hbar \equiv 1$)
\begin{eqnarray}
H_k &=& \int d^{3}\mathbf{r} \Psi_k^{\dag} \left[
-\frac{\nabla^2}{2m} + V_k + \sum_l
\frac{U_{kl}}{2} \Psi_k^{\dag} \Psi_k \right] \Psi_k, \\
H_{\rm{ext}} &=& \frac{1}{2} \int d^{3}\mathbf r \left[
\Psi_A^{\dag} \Psi_B \Omega_{\rm{R}} e^{-i \Delta t} +
\Psi_B^{\dag} \Psi_A \Omega_{\rm{R}}^{\ast}e^{+i \Delta t}
\right]. \nonumber \label{FullLaserCoupling}
\end{eqnarray}
Here $\Psi_k \equiv \Psi_k(\mathbf{r})$ ($k = \{A,B\}$) are
bosonic field operators that annihilate a particle at position
$\mathbf{r}$ in the hyperfine state $\st{k}$. The trapping
potential for atoms in state $k$ is denoted by $V_k \equiv
V_k(\mathbf{r})$ and their mass is $m$. The interaction strengths
are given by $U_{AA}, U_{BB}$ and $U_{AB}$ for collisions between
particles in state $A$, $B$ and interspecies collisions,
respectively. The effective Rabi frequency $\Omega_{\rm{R}}$ is
assumed to be positive, real, and position independent, i.e., we
neglect the momentum transfer induced by the external field. The
detuning of the field from resonance is denoted by $\Delta$.

\subsubsection{Two mode approximation}
\label{sec:TMA}

We assume that the spatial degrees of freedom can be described
using one spatial mode function for each component
\cite{Cirac97,Sorensen01,Duan}
\begin{equation}
\Psi_A(\mathbf{r})=a \psi_A(\mathbf{r}) \mathand
\Psi_B(\mathbf{r})=b \psi_B(\mathbf{r}), \label{TwoModeAnsatz}
\end{equation}
where $\psi_k(\mathbf{r})$ are real normalized wave functions and
$a$, $(b)$ are bosonic annihilation operators destroying a particle
in the internal state $A$, $(B)$. They obey the usual bosonic
commutation relations $[a,a^\dagger]=1$, $[b,b^\dagger]=1$,
$[a,b]=0$ and $[a,b^\dagger]=0$.

\subsubsection{Angular momentum representation}

We use the angular momentum operators $\bf{S}=(S_x,S_y,S_z)$
defined in Eq.~(\ref{AngularMomentumTwoMode}) and the eigenstates
of $S_z$ with eigenvalue $n$:
\begin{equation}
\st{n}_z =
\frac{\left(a^{\dag}\right)^{N/2+n}\left(b^{\dag}\right)^{N/2-n}}{\sqrt{\left(N/2+n\right)!\left(N/2-n\right)!}}
\st{\rm{vac}},
\end{equation}
The operators $\mathbf{S}$ fulfill the standard angular momentum
commutation relations $\left[S_i,S_j\right]=i S_k$, with $i,j,k$
cyclic. From the Heisenberg uncertainty principle we find
$\Av{\Delta S_i^2} \Av{\Delta S_j^2} \geq \frac{1}{4}
|\Av{S_k}|^2$.

\subsubsection{Two mode Hamiltonian}\label{TMHam}

Using Eqs.~(\ref{AngularMomentumTwoMode},\ref{TwoModeAnsatz}) the
Hamiltonian reduces (up to a constant) to
\begin{equation}
H = \delta S_z + \chi S_z^2 + \Omega S_x,
\label{HamiltonianAngularMomentum}
\end{equation}
where $\delta = \omega_A - \omega_B + \Delta + (u_{AA}-u_{BB})(N -
1)/2$, and $\chi = (u_{AA}+u_{BB} - 2 u_{AB})/2$ with . The single
particle ground state energies $\omega_A$ and $\omega_B$, the
coupling $u_{kl}$ and the effective Rabi frequency $\Omega$ are
given by
\begin{eqnarray}
\omega_k &=& \int d^{3}\mathbf r \psi_k^{*} \left[
-\frac{\nabla^2}{2m}
+ V_k \right] \psi_k, \nonumber \\
u_{kl} &=& U_{kl} \int d^{3}\mathbf r |\psi_k|^2 |\psi_l|^2, \nonumber  \\
\Omega &=& \Omega_{\rm{R}} \int d^{3}\mathbf r \psi_A^{*} \psi_B.
\end{eqnarray}
In the following we will consider the situation, where $\chi>0$
and $\delta=0$.

\subsection{Phase model}
\label{PhaseModel1}

In the introduction we used the phase model
\cite{Menotti1,Jaksch1} to qualitatively explain how the time
evolution according to $H$ can be deployed to create many particle
entangled states. To be more precise the continuous eigenstates of
$\phi$ are given by
\begin{equation}
\st{\phi}=\frac{1}{\sqrt{2 \pi}} \sum_{n=-N/2}^{+N/2} e^{-i n
\phi} \st{n}_z.
\end{equation}
Using the relation $S_z \st{\phi} = -i \partial_{\phi} \st{\phi}$,
neglecting terms of order $1/N$ and for $\Omega \ll N \chi / 2$ we
can rewrite the Hamiltonian $H$ \cite{Menotti1} in the phase
representation as given in Eq.~(\ref{HamiltonPhase}). The
Hamiltonian $H_\phi$ describes a single fictitious particle moving
as a pendulum \cite{Leggett1,Smerzi1}. The first term of $H_\phi$
can easily be identified as the kinetic energy of the particle and
the second term is a conservative periodic potential. For large $N
\gg 1$ the initial phase wave function $\Psi(0,\phi) \equiv \langle
\phi \st{\Psi (t=0)}$ is well approximated by a narrow gaussian
wave-packet
\begin{eqnarray}
\Psi(0, \phi)&=& \frac{1}{\sqrt{2^{N+1} \pi}} \sum_{n=-N/2}^{+N/2}
\left(\begin{array}{c} N \\
N/2-n \end{array} \right)^{1/2} e^{i n \phi} \nonumber \\ &\approx&
\left( 2 \pi \sigma^2 \right)^{-1/4} e^{-\frac{\phi^2}{4
\sigma^2}}.
\end{eqnarray}
The width of the Gaussian is $\sigma=1/\sqrt{N}$. The phase model
is valid for $\Omega \ll \chi N$ \cite{Menotti1}. Since we do not
want to restrict ourselves to this case we will now derive a
semiclassical model valid for arbitrary $\Omega$.

\subsection{Semiclassical model}
\label{SemiClassical}

The many particle entangled state we are interested in are
superposition states of two wave packets centered at two different
relative particle numbers. We want to use a semiclassical model to
estimate the position of the maxima of these wave packets. We
assume the collective spin $\bf S$ to behave like a classical
quantity, which is possible if the discreteness of the energy
levels is negligible i.e. for $N \gg 1$. Furthermore the
semiclassical treatment will only be valid as long as interference
effects are negligible. For the initial state we consider it will
turn out that such interference effects become important for times
$t > 2 t_{\rm{c}}$, where $t_{\rm{c}}$ defined in
Eq.~\eqref{eq:timescale} is the time it takes to create the many
particle entangled state.

\begin{figure*}[tb]
\begin{center}
\epsffile{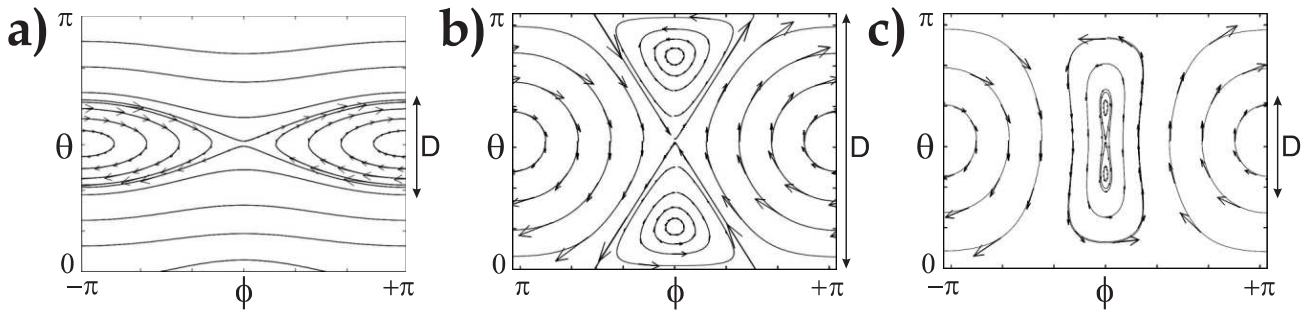}
\caption{Vector field (arrows) and trajectories (solid curves) of
Eq.~\eqref{classicalVF} for three different coupling strengths: a)
$\omega=0.14$, b) $\omega=1$ and c) $\omega=1.87$, where
$\omega\equiv 2 \Omega / \chi N$.} \label{fig:fig_vf_field}
\end{center}
\end{figure*}

Under these conditions we replace the spin operator $\mathbf{S}$
by c-numbers
\begin{equation}
\mathbf{S} \rightarrow \frac N 2 \left( \sin \theta \cos \phi ,
\sin \theta \sin \phi , \cos \theta \right).
\end{equation}
This implies the factorization of expectation values of products
of operators like e.g. $\Av{ \{ S_x , S_y \} }$ by $2 \Av{S_x}
\Av{S_y}$. From the Heisenberg equations of motion for the
operator $\mathbf S$ given by
\begin{equation}
\frac{d}{dt} \left(
\begin{array}{c}
S_x \\ S_y \\ S_z
\end{array} \right) =
\left( \begin{array}{c} - \chi \{ S_z, S_y \}
\\ \chi \{ S_z, S_x \} -\Omega S_z
\\ \Omega S_y
\end{array} \right)
\end{equation}
we obtain for the time evolution of the angles $\theta, \phi$
\begin{eqnarray}
\frac{d}{dt} \left(
\begin{array}{c} \theta
\\ \phi \end{array} \right) = \frac{\chi N}{2} \left( \begin{array}{c} -\omega \sin \phi  \\
2 \cos \theta - \omega \cot \theta \cos \phi
\end{array} \right),
\label{classicalVF}
\end{eqnarray}
with $\omega\equiv 2 \Omega / \chi N$. The corresponding vector
fields are shown in Fig.~\ref{fig:fig_vf_field} for different
values of $\omega$.

From the continuous Wigner function $W_t$ \cite{Leonhardt} defined
as the Fourier transform of the quantum characteristic function
$\tilde W_t$ given by
\begin{eqnarray}
W_t(n,\phi) &=& (2 \pi)^{-2} \int d \phi' \int d n' \tilde W_t(n',\phi') e^{+i (\phi' n - n' \phi)}, \nonumber \\
\tilde W_t(n',\phi') &=& \stdag{\Psi(t)} e^{i \left( n' \hat \phi -
\phi' S_z \right)} \st{\Psi(t)}, \label{eq:WignerCont}
\end{eqnarray}
we get the initial Liouville distribution on the sphere as
\begin{equation}
P_{t=0}(\theta,\phi) = \frac{1}{\sin \theta} W_{t=0}(n=S \cos \theta,\phi).
\end{equation}
For the initial state $\st{\Psi(t=0)}$ the  Liouville distribution
is well approximated by a Gaussian of narrow radial width
$\sigma=1/\sqrt{N}$:
\begin{equation}
\begin{split}
P_0(\theta,\phi) = \mathcal{N} e^{-\frac{2}{\sigma^2 N^2}(\mathbf S
- \mathbf S_0)^2} \approx
\frac{1}{2\pi\sigma^2}e^{-\frac{1}{2 \sigma^2}(\cos^2
\theta+\phi^2)}.
\end{split}
\label{eq:liouvillePsi0Sx}
\end{equation}
The semiclassical time evolution of the Liouville distribution is
then given by
\begin{equation}
\begin{split}
P_t(\theta,\phi)=\frac{1}{\sin \theta} \int d\theta' \sin \theta' \int d\phi' P_0(\theta',\phi')\times \\
\delta(\theta-\tilde \theta_t(\theta',\phi')) \delta(\phi - \tilde \phi_t(\theta',\phi')),
\end{split}
\label{eq:liouville}
\end{equation}
where $\tilde \theta_t(\theta',\phi'), \tilde
\phi_t(\theta',\phi')$ are the classical trajectories of a
fictional particle starting at $t=0$ from $\theta',\phi'$, i.e.
solutions of Eq.~\eqref{classicalVF}. The semiclassical
expectation value of an operator $C({\bf S})$ is given by
\begin{eqnarray}
\Av{C({\bf S})}_t&=&\int_{0}^{\pi} d\theta \sin \theta
\int_{-\pi}^{+\pi} d\phi P_t(\theta,\phi) \nonumber \\
&&C\left[\frac N 2 \left( \sin \theta \cos \phi , \sin \theta \sin
\phi , \cos \theta \right)\right].
\end{eqnarray}

\section{Engineering many-particle entangled states}
\label{Overview}

In this section we first define different kinds of many particle
entangled states and review several schemes for engineering
entanglement in two component BECs. We then study in detail a
scheme which allows the production of a many particle entangled
state in a two component BEC at a time scale proportional to
$1/N$.

\subsection{Many particle states}

We introduce different kinds of many particle states which will be
used to characterize the states encountered in the numerical
calculations performed in the subsequent section \ref{sec:numres}.

\subsubsection{Coherent spin states (CSS)}

The CSS \cite{Perelomov} are eigenstates of the angular momentum
operator $S_{\mathbf n_1} \equiv \mathbf{n_1}\cdot\mathbf{S}$ with
eigenvalue $N/2$, where $\mathbf{n_1}$ is a unit vector pointing in
the direction $(\theta,\phi)$. These states are completely
uncorrelated and can be written as a separable product of
single-particle states given by
\begin{eqnarray}
\st{\theta,\phi} &\equiv&  e^{-i S_z \phi} e^{-i S_y \theta}
\st{+N/2}_z \nonumber \\
&=&  \frac{1}{\sqrt{N!}}\left[\cos \frac{\theta}{2} e^{-i\phi/2}
a^\dag + \sin
\frac{\theta}{2} e^{+i\phi/2}b^\dag\right]^N\st{\rm{vac}} \nonumber \\
&=& \left[\cos \frac{\theta}{2} e^{-i\phi/2}\st{A}+\sin
\frac{\theta}{2} e^{+i\phi/2}\st{B}\right]^{\otimes{N}}.
\end{eqnarray}
CSS are minimum uncertainty states with $\Av{\Delta S_{\mathbf
n_3}^2} = \Av{\Delta S_{\mathbf n_3}^2} = N/4$, where ${\mathbf
n_1}, {\mathbf n_2}, {\mathbf n_3}$, are unit vectors orthogonal
to each other.

\subsubsection{Spin squeezed states (SSS)}

The SSS \cite{Kitagawa1,Heinzen1} are characterized by a reduced
variance compared to that of the CSS in one of the spin-components
$S_{\mathbf n_2}$ or $S_{\mathbf n_3}$ whereas the variance in the
other orthogonal component is correspondingly enhanced. The amount
of squeezing is determined by the squeezing parameter $\xi$ given
by
\begin{equation}
\xi^2=\min_{\mathbf{n}_{1,2,3}} \frac{N (\Delta
S_{\mathbf{n_1}})^{2}}{\Av{S_{\mathbf{n_2}}}^{2}+\Av{S_{\mathbf{n_3}}}^{2}},
\end{equation}
If $\xi^2 < 1$ the state of the atoms is non--separable (i.e.
entangled) as has been shown in \cite{Sorensen01}. The parameter
$\xi^2$ thus characterizes the atomic entanglement, and we refer to
states with $\xi^2 < 1$ as ``spin squeezed states''.

We also note that this parameter determines the amount of
noise-reduction in atomic clocks \cite{Heinzen1}. Therefore the
robust creation \cite{Sorensen01} and preservation \cite{Jaksch1}
of SSS might prove useful in enhancing the accuracy of these
atomic clocks.

\subsubsection{Maximally entangled states (MES)}

A second type of macroscopically entangled states can be written
as
\begin{eqnarray}
\st{{\rm GHZ}}_N &=& \frac{1}{\sqrt{2}} \left( \st{A}^{\otimes N} +
\st{B}^{\otimes N} \right)  \nonumber \\
&=& \frac{1}{\sqrt{2}} \left( \st{+N/2}_z + \st{-N/2}_z \right)
\nonumber \\
&=& \frac{1}{\sqrt{2}} \left( \st{\theta=0,\phi=0} +
\st{\theta=\pi,\phi=0} \right). \label{GHZstate}
\end{eqnarray}
They are a generalization for $N$ particles of the well known
GHZ-state of three particles
\cite{Bouwmeester,Bennett1Measures,GHZreview}. As can be seen from
Eq.~(\ref{GHZstate}) these states can be written as either a
coherent superposition of $N$ particles being in the state $A$ and
$N$ being in state $B$ or as a coherent superposition of two CSS
pointing in opposite directions on the Bloch-sphere.

\subsection{Schemes for producing many particle entanglement}

There have been several proposals of how to engineer many particle
entanglement in BECs. Basically these schemes can be divided into
two different kinds. The first possibility is to engineer the
ground state of the system to be an entangled state. This can be
done by appropriately manipulating the interaction between the
particles \cite{Cirac97,Search1}. The entangled state is then
created by cooling the system to its ground state. The second
possibility is to control the dynamics of the system such that an
initially separable state evolves coherently into an entangled
state \cite{SorensenHot,Helmerson1,Duan,GordonSavageCat}.

\subsubsection{Thermodynamical Schemes}

For $\chi>0$ and $\Omega=0$ the ground state of the Hamiltonian
\eqref{Hamiltonian} is the number-squeezed (``dual fock'')
\cite{Kasevich} state $\st{0}_z$, with half of the atoms in
$\st{A}$ and the others in $\st{B}$. As first noted by Cirac
\textit{et al.} \cite{Cirac97} for $\chi<0$ and $|\Omega|< N/2$ the
ground state of the system corresponds to an entangled state and a
maximally entangled state of the form Eq.~(\ref{GHZstate}) is
attained for zero coupling (i.e.~$\Omega \rightarrow 0$). The major
obstacle in cooling to the ground state is the small energy gap
between the ground and the first excited state which scales like
$\chi$.

\subsubsection{Dynamical Schemes}

One of the simplest ways \cite{GordonSavageCat} to obtain a SSS is
to start from a one component BEC in $A$ (corresponding to a CSS
$\st{N/2}_{-z}$), to apply a fast $\pi/2$ pulse ($\Omega \gg \chi
N$) that rotates the CSS on the Bloch sphere by an angle $\pi/2$
around the $y$-axis aligning it along the $\mathbf x$ direction
(corresponding to $\st{N/2}_{x}$). The subsequent evolution due to
the Hamiltonian $H$ (with $\Omega=0$) establishes correlations
among the particles, creating an SSS with a squeezing parameter
$\xi \approx (3/N)^{2/3}/2$ for large $N$ \cite{Kitagawa1} on a
time scale $t \approx 2 \times 3^{1/6}/\chi N^{2/3}$. Taking into
account the spatial degrees of freedom and inelastic collisions
with background particles still allows for considerable squeezing
of the SSS as was more recently shown in \cite{Sorensen01,
SorensenBogoliubov1, Poulsen1}. As noted by M{\o}lmer {\it et al.}
\cite{SorensenHot} and Castin \cite{CastinLesHouches} one-axis
squeezing also provides a perfect GHZ-state at a much later time
$t=\pi/2 \chi$ neglecting, however, the spatial degrees of freedom
and decoherence processes.

Even smaller squeezing parameters $\xi$ can be obtained by
engineering two particle interactions resulting in a Hamiltonian of
the form $H_{\rm int} =\chi (S_z^2-S_y^2)$ as recently proposed in
\cite{Helmerson1, Duan} and thus implementing two-axis-squeezing
\cite{Kitagawa1}. In \cite{Helmerson1} intermediate molecular
states of two-atoms are used to create a Hamiltonian of the form
$H_{\rm int}$ while in \cite{Duan} this is achieved by applying a
series of laser pulses to the two-component BEC. As discussed by
Law {\it et al.} \cite{Law2} turning on a small coupling $\Omega
\ll \chi N$ can be used to further improve the squeezing
properties. Gordon {\it et al.} \cite{GordonSavageCat} showed
numerically that for $\Omega \sim \chi N$ Schr\"{o}dinger Cat
states can be obtained on short time scales.

The main task of the remainder of this paper is to investigate in
detail dynamical schemes which create many particle entangled
states of the form
\begin{eqnarray}
\st{\rm{Cat}(D)}_N&=&\frac{1}{\sqrt{2}}\left(\st{+\frac{D}{2}}_z
+\st{-\frac{D}{2}}_z \right), \label{CatState1} \\
\st{\rm{Cat}(\gamma)}_N&=&\frac{1}{\sqrt{2}}\left(\st{\frac{\pi+
\gamma}{2},0}+\st{\frac{\pi-\gamma}{2},0}\right),
\label{CatState2}
\end{eqnarray}
on a time scale proportional to $1/N$ using the Hamiltonian
$H_{\rm BEC}$ (see also \cite{GordonSavageCat}).
Eq.~(\ref{CatState1}) is a macroscopic superposition of
eigenstates of the operator $S_z$ characterized by a ``distance''
$D$. The other entangled state Eq.~(\ref{CatState2}) is a
superposition of two CSS separated by an angle $\gamma$. Note that
for $D=0$ and $\gamma=0$ these states are not entangled, whereas
for $D=N$ and $\gamma=\pi$ they coincide forming an MES.

\section{Semiclassical Results} \label{sec:SCRes}

In this section we first identify three different coupling
regimes, for which the evolution of the initial state has a
qualitatively different behavior. Then we derive analytical
approximations for the time scales on which the many particle
entangled states Eq.~(\ref{CatState1}) are formed, and the
attainable separation $D$ as a function of $\Omega$. We find that
for $\Omega=\chi N/2$ states close to MES are created on a time
scale $t_c = \ln(8N)/\chi N$.

\subsection{Coupling regimes} \label{sec:regimes}

As can be seen from Fig.~\ref{fig:fig_vf_field} the vector field
\eqref{classicalVF} on the Bloch-sphere is qualitatively different
for $2 \Omega / \chi N \equiv \omega < 1$ (weak coupling regime)
and $\omega > 1$ (strong coupling regime). We will thus
investigate these two regimes separately and also look at the
intermediate case $\omega = 1$ (critical coupling).

In the weak coupling regime the trajectories passing through the
maximum of our initial distribution at $\theta=\pi/2, \phi=0$,
make a full revolution in $\phi$. The separatrix separates the
rotational modes of the pendulum from the oscillatory ones. The
rotational modes of the pendulum rotate continuously either
clockwise- or counter-clockwise (trajectories in the lower/upper
half plane of $\theta$ outside the separatrix), whereas the
oscillatory modes oscillate around $\phi=-\pi/2$ (trajectories
inside the separatrix), as shown in Fig.~\ref{fig:fig_vf_field}a.
In this regime the time evolution first squeezes our initial state
along the separatrix. Then two elongated peaks aligned along the
$\phi$-axis centered at $\phi=\pm \pi$ appear, respectively (see
Fig.~\ref{fig:ComparisonLow}a2-b2). From the reduced density
distribution
\begin{eqnarray}
P_{\rm{sc}}(n,t) &=& \int d\phi \tilde P_t(n,\phi),\\
\tilde P_t(n=\cos \theta,\phi) &=& \sin \theta P_t(\cos
\theta,\phi),
\end{eqnarray}
we find good agreement with the Wigner distribution for a state of
the form Eq.~\eqref{CatState1} as shown in
Fig.~\ref{fig:ComparisonLow}a4-b4.

In the strong coupling regime the trajectories passing through
$\theta=\pi/2, \phi=0$, do neither perform a full revolution in
$\phi$ nor separate the rotational from the oscillatory modes any
more as can be seen from Fig.~\ref{fig:fig_vf_field}c. In fact,
there are no rotational modes for $\Omega
> \chi S$. However, first the initial state again is squeezed like
in the weak coupling regime (see
Fig.~\ref{fig:ComparisonHigh}a1,b1). Then two well separated
gaussian peaks - one on the northern and another at the southern
hemisphere appear. In contrast to the weak coupling regime they
are now centered at $\phi=0$ as shown in
Fig.~\ref{fig:ComparisonHigh}a2,b2. This distribution corresponds
roughly to a state of the form Eq.~\eqref{CatState2} as shown in
Fig.~\ref{fig:ComparisonHigh}b4.

\subsection{Distance}
\label{sec:Distance}

According to the semiclassical time evolution the largest distance
$D$ of the many particle entangled state is equivalent to the
largest separation of the separatrix at $\phi=\pi$ in the weak
coupling regime and at $\phi=0$ in the strong coupling regime. By
using the conservation of energy $E(\theta,\phi)=\chi (N \cos
(\theta)/2)^2 + \Omega N \sin (\theta) \cos (\phi)/2 = \Omega
N/2$, we find the shape of the separatrix
\begin{equation}
\begin{split}
\cos \phi[\theta]=\frac{1-\omega^{-1} \cos^2 \theta}{\sin \theta}
 \quad
\Leftrightarrow \\
\sin \theta[\phi]=\frac{\omega}{2} \cos \phi \pm
\sqrt{ \left( \frac{\omega}{2} \cos \phi \right)^2 - \omega + 1}.
\end{split}
\label{eq:separatrix}
\end{equation}
It follows that in the weak coupling regime the distance $D$ is
given by (cf. Appendix \ref{AppendixB})
\begin{equation}
D=N \sqrt{\omega (2-\omega)}, \label{eq:distance}
\end{equation}
and in the strong coupling regime the angle between the two CSS
states $\gamma$ is
\begin{equation}
\gamma=2 \arcsin(\sqrt{\omega (2-\omega)}). \label{eq:angle}
\end{equation}

In the weak coupling regime the maximum distance $D$ increases with
$\omega$ until $\omega=1$ where $D$ takes its largest possible
value $D=N$ (see Fig.~\ref{fig:DistanceAndMore}a). Thus for
$\omega=1$ we obtain a state close to a MES. For larger couplings
($\omega>1$) a superposition state of the form
Eq.~\eqref{CatState2} is created. The angle $\gamma$ decreases with
increasing $\omega$ until at $\omega=2$ no superposition is
obtained according to Eq.~\eqref{eq:angle}. Exact numerical
calculations show that for $\omega \geq 2$ macroscopically
entangled states of the latter form are still obtained but with
$\gamma \ll \pi$.

\subsection{Time Scales}

We will now focus on the required time $t_{\rm{c}}$ to create a
many particle entangled state by using the semiclassical time
evolution of the wave packets forming the superposition states. It
is crucial that $t_c$ is short compared to decoherence times to
successfully create many particle entanglement.

The time $t_{\rm{c}}$ is approximately equal to the time needed to
``travel'' along the separatrix from the point lying at the
distance $\sigma$ from $\phi=0, \theta=\pi/2$ to $\cos \theta=D/N$,
i.e., the point where the many particle entangled state forms.
Combining Eqs.~(\ref{classicalVF},\ref{eq:separatrix},
\ref{eq:distance}) one obtains (cf. Appendix \ref{AppendixB})
\begin{equation}
\begin{split}
\chi t_{\rm{c}} = & \frac{2 \log\left[\sqrt{2N(2-\omega)}+\sqrt{2N(2-\omega)-1}\right]}{N \sqrt{\omega(2-\omega)}} \\
& \approx
\frac{\log\left[8N(2-\omega)\right]}{N\sqrt{\omega(2-\omega)}},
\end{split}
\label{eq:timescale}
\end{equation}
where the last approximation is valid for $2 N (2-\omega) \gg 1$.
We notice that $t_c$ scales as $1/N$ and reaches its minimum value
for  $\omega \approx 1$, i.e., near the critical coupling, see
Fig.~\ref{fig:DistanceAndMore}b.

\section{Numerical Results}\label{sec:numres}

We calculated numerically the solution of the Schr\"{o}dinger
equation for the exact model
Eq.~\eqref{HamiltonianAngularMomentum} with a moderate number of
particles $N$ up to $10^3$. A comparison of the exact results with
the approximations of Sec.~\ref{sec:SCRes} yields good agreement
already for $N \geq 50$. Also, we calculate the overlap of the
cat-states $\st{\psi}$ with the states defined in
Eqs.~(\ref{CatState1},\ref{CatState2}), i.e., their fidelity
\begin{equation}
F_D=|{}_N\stdag{{\rm Cat}(D)} \psi \rangle |^2.
\end{equation}
Furthermore we find a partial revival of the initial wave function
at time $t \sim 2 t_c$. We characterize this revival by the overlap
with the initial CSS state $R$ given by
\begin{equation}
R = |\langle \Psi(0) \st{\Psi(t)}|^2 = |{}_x\langle N/2
\st{\Psi(t)}|^2.
\end{equation}
Since $R$ turns out to be larger than $1/2$  for $0.1<\omega<1.9$
it is useful for measurement purposes, especially for checking the
coherence of an MES, as described in detail in
Sec.(\ref{sec:measurement}).

\subsection{Comparison of Semiclassical with Exact Results} \label{sec:Comparison}

\begin{figure}[ht]
\begin{center}
\epsffile{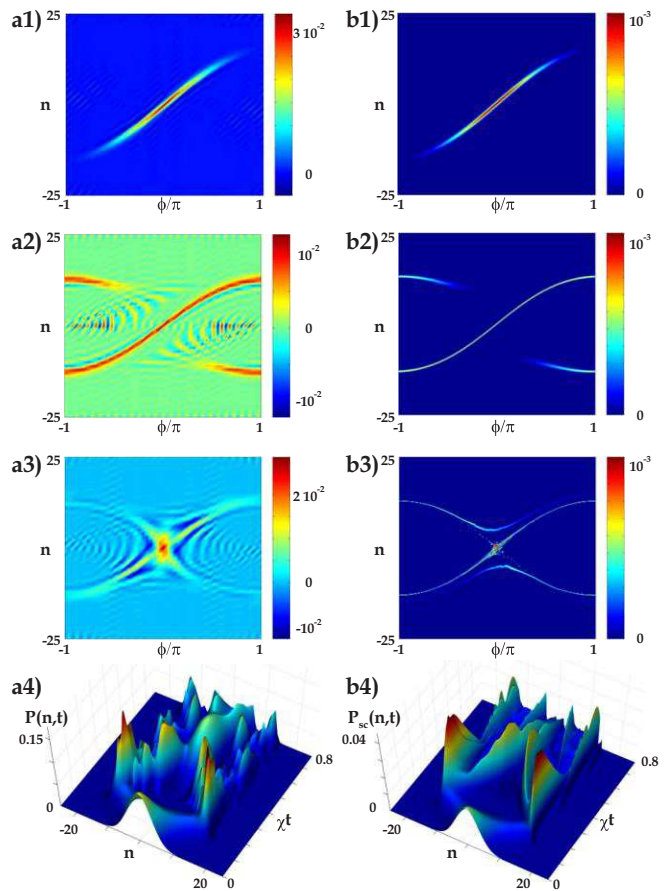}
\end{center}
\caption{Comparison of the evolution of the system for
$\omega=0.14$ (weak coupling) and $N=50$. The exact discrete Wigner
function $W_t(n,\phi)$ is plotted in (a1-a3) compared with the
Lioville distribution $\tilde P_t(n=\cos \theta,\phi)$ in (b1-b3)
at three characteristic times. 1) $t=t_c/2$ showing a SSS; 2)
$t=t_c$ showing the macroscopic superpositon state; and in 3) $t
\sim 2 t_c$ showing the revival. a4) exact discrete number
distribution $P(n,t) \equiv |{}_z\langle n \st{\Psi(t)}|^2$; b4)
reduced density distribution $P_{\rm{sc}}(n,t)$.}
\label{fig:ComparisonLow}
\end{figure}

\begin{figure}[ht]
\begin{center}
\epsffile{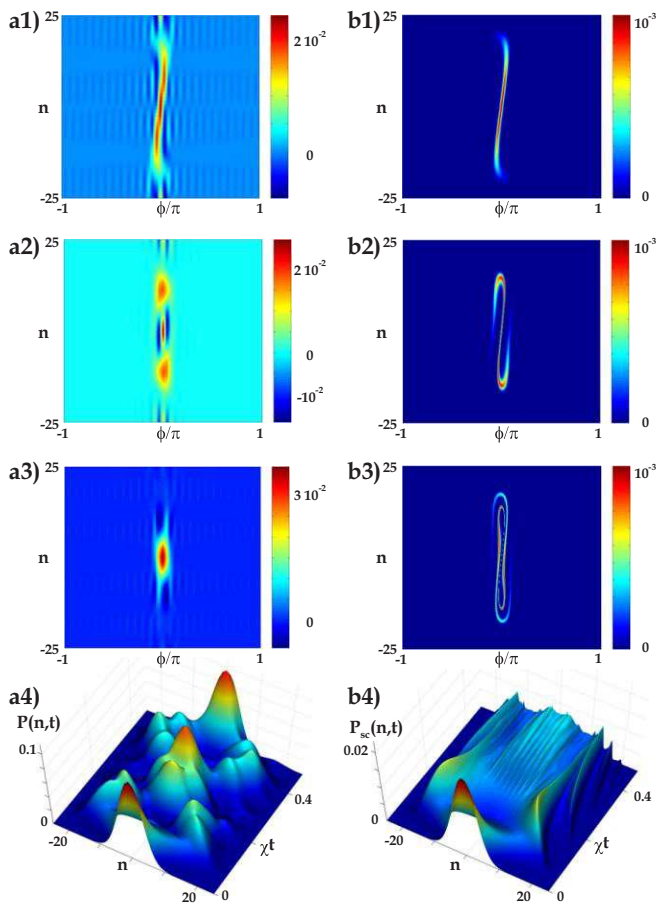}
\end{center}
\caption{Same as Fig.~\ref{fig:ComparisonLow} but for
$\omega=1.87$ (strong coupling).} \label{fig:ComparisonHigh}
\end{figure}

We calculate numerically the solution of the Liouville equation
Eq.~\eqref{eq:liouville} and compare it with the exact discrete
Wigner function as defined by Leonhardt \cite{Leonhardt} for
$0<\omega<2$. From Figs.~\ref{fig:ComparisonLow}a1-b2,
\ref{fig:ComparisonHigh}a1-b2 we see excellent agreement between
the two results for $N \geq 50$ and $t<2 t_c$. Afterwards
interference effects become important and the semiclassical model
breaks down as shown in Figs.~ \ref{fig:ComparisonLow}a3-b3,
\ref{fig:ComparisonHigh}a3-b3. As can be seen by comparing Figs.~
\ref{fig:ComparisonLow}a4-b4 and Figs.~\ref{fig:ComparisonLow}a4-b4
we obtain a revival of the initial state of about $55{\%}$ [$90
\%$] for the exact solution in the weak [strong] coupling regime,
whereas the overlap of the semiclassical solution with the initial
state is only about $10{\%}$.

\begin{figure}[ht]
\begin{center}
\epsffile{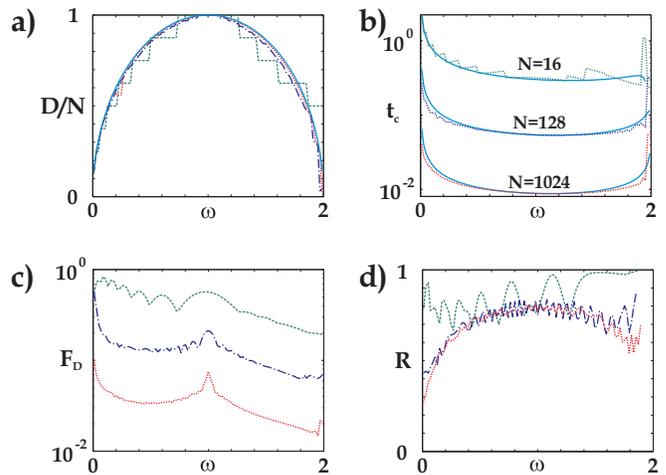}
\end{center}
\caption{Comparison of the exact with the semiclassical results. a)
Attainable distance $D$ as a function of $\omega$, exact solution
for $N=16$ (dashed curve), $N=128$ (dash-dotted curve), $N=1024$
(dotted curve), and semiclassical result (solid curve). b) Time
$t_c$ as a function of $\omega$: exact solution (dashed curves)
compared with the semiclassical result (solid curves) for
$N=16,128,1024$. c) The maximum fidelity $F_D$ as a function of
$\omega$. d) Revival $R$ for $N=16$ (dashed curve), $N=128$
(dash-dotted curve), and $N=512$ (dotted curve).}
\label{fig:DistanceAndMore}
\end{figure}

Next we proceed to check the validity of
Eqs.~(\ref{eq:distance},\ref{eq:timescale}) for $D$ and $t_c$.
Numerically we find $D$ and $t_c$ by looking for the maximum
$F_D(t) > F_0(t)$. The numerical values for $D$ and $t_c$ agree
very well with the analytical expressions
Eqs.~(\ref{eq:distance},\ref{eq:timescale}) for $N \geq 50$, see
Fig.~\ref{fig:DistanceAndMore}a,b.

\begin{figure}[ht]
\begin{center}
\epsffile{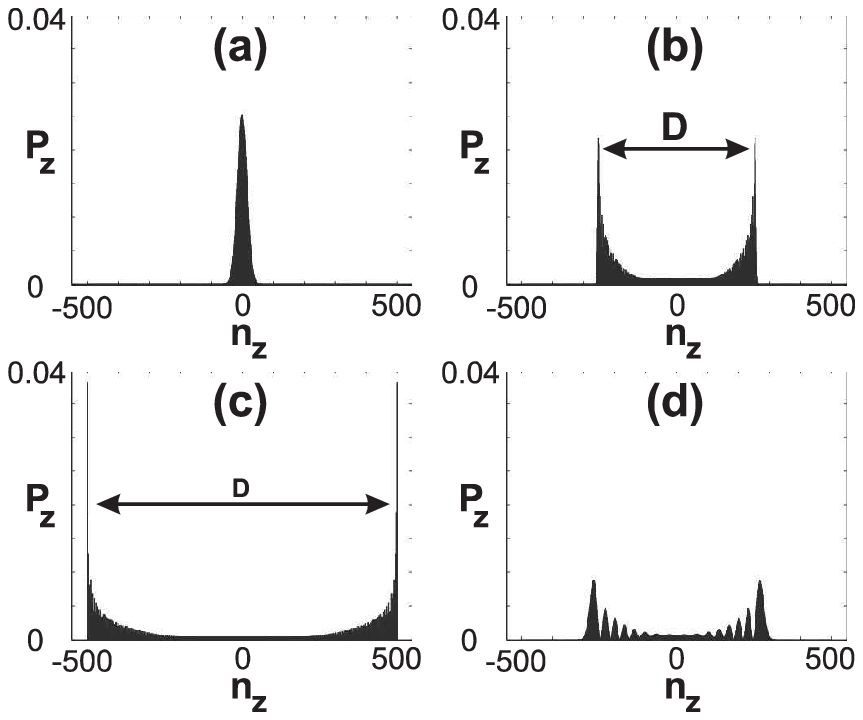}
\end{center}
\caption{Exact probability distribution $P(n,t)$ for $N=10^3$
particles for (a) the initial state, (b) the state created for
$\omega \approx 0.14$ corresponding to a distance of $D=N/2$, (c)
the GHZ-like state obtained for $\omega=1$, and (d) the state for
$\omega \approx 1.87$ corresponding to an angle of $\gamma=2
\arcsin (1/2)$.} \label{fig_bar_pics}
\end{figure}

\subsection{Critical Coupling} \label{sec:CriticalCoupling}

\begin{figure}[tbp]
\begin{center}
\epsffile{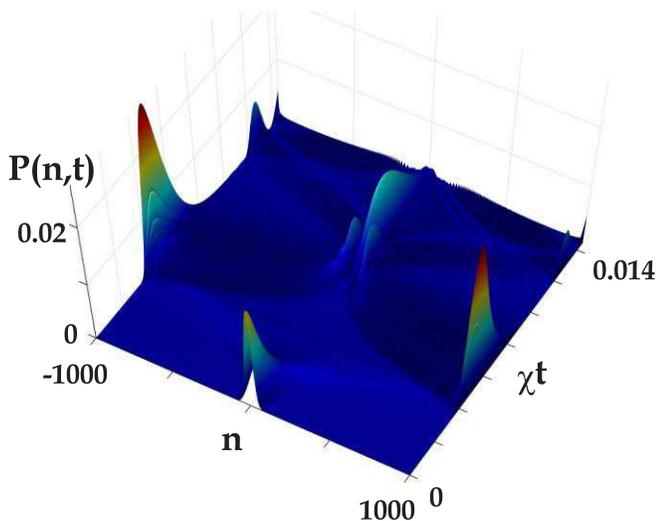}
\end{center}
\caption{Exact number distribution $P(n,t)$ as a function of $n$
and $t$ for $\omega=1$ and $N=2000$.} \label{fig:evolutionN2000CC}
\end{figure}

We see from Fig.~\ref{fig:DistanceAndMore} that for the critical
coupling $\omega=1$ we obtain the largest possible $D=N$, a short
creation time $t_c$, the best fidelity $F_D$, and the largest value
for the partial revival $R$. Also the decrease of $F_D$ with
increasing number of particles $N$ is slowest at the critical
coupling. In Fig.~\ref{fig:evolutionN2000CC} we plot the time
evolution showing the initial squeezing followed by the creation of
a state close to MES and the partial revival of the initial state
with $R=79{\%}$.

\section{Pre-Squeezing}\label{sec:One-Axis-Pre-Squeezing}

\begin{figure}[ht]
\begin{center}
\epsffile{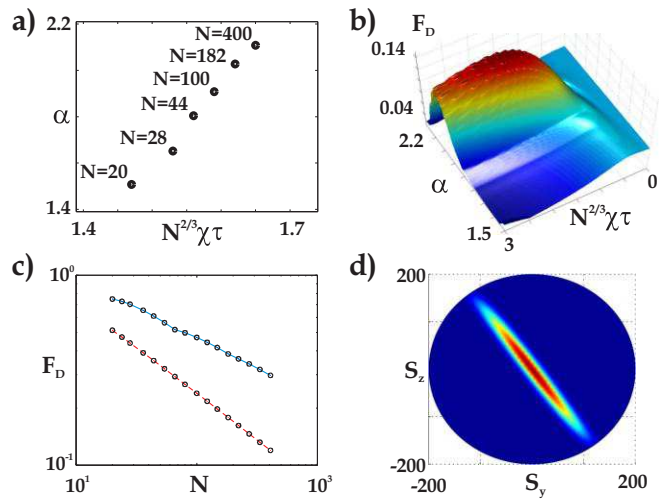}
\end{center}
\caption{a) Optimum values for $\tau$, $\alpha$ for different
values of $N$. b) Fidelity $F_D$ as a function of $\tau$ and
$\alpha$ for $N=400$. c) Comparison of $F_D$ without pre-squeezing
(dash-dotted curve), and with pre-squeezing (solid curve) as a
function of $N$, where the dots mark the numerically calculated
values. d) Initial CSS-Q-function $Q(\theta,\phi)\equiv|\langle
\theta,\phi \st{\Psi}|^2$ for optimal pre-squeezing for $N=400$.}
\label{fig:1axispresqueezed}
\end{figure}

We investigate if better results for $F_D$ are obtainable by
adjusting the initial variances $\Delta S_y, \Delta S_z$ by
squeezing. We choose one-axis-squeezing, which is easily
implemented by turning off the external field after the initial
$\pi/2$-pulse for a time $\tau$. Afterwards a second pulse with
angle $\alpha$ is applied before the system is evolved according
to Eq.~\eqref{HamiltonianAngularMomentum}. We optimize $\tau$ and
$\alpha$ for obtaining the largest $F_D$ and restrict ourselves to
the most interesting case $\omega=1$. The results are shown in
Fig.~\ref{fig:1axispresqueezed}. A substantial improvement in the
obtainable fidelity $F_D$ can thus be reached by pre-squeezing,
(cf.~Fig.~\ref{fig:1axispresqueezed}c) which, furthermore,
decreases much slower with increasing $N$ than without
pre-squeezing.

\section{Discussion}\label{Disc}

\subsection{Stability under Decoherence}
\label{sec:decoherence}

A major problem of creating a macroscopic entangled state is
decoherence, affecting every realistic system. For instance the
entanglement in an MES is destroyed by losing a single particle,
as is easily shown by calculating
\begin{eqnarray}
\st{\tilde \Psi_A}_{\tilde N} &= \frac{a \st{\rm{GHZ}}_N}{|| a
\st{\rm{GHZ}}_N||} &=
\st{A}^{\otimes \tilde N} = \st{\tilde N : \tilde N/2}_z, \\
\st{\tilde \Psi_B}_{\tilde N} &= \frac{b \st{\rm{GHZ}}_N}{|| b
\st{\rm{GHZ}}_N||} &= \st{B}^{\otimes \tilde N} = \st{\tilde N :
-\tilde N/2}_z,
\end{eqnarray}
with $\tilde N = N -1$ and $\st{N: n}_z$ the eigenstate of $S_z$
for $N$ particles.

In our scheme, however, we expect the effects of decoherence to be
reduced for two reasons. (i) The many particle entangled state is
created on a very short time scale, for which we expect only few
particles to be lost. (ii) Losing a single particle does not
completely destroy the entanglement for $D<N$, since
\begin{eqnarray}
 \st{\tilde \Psi_A}_{\tilde N} =& \frac{a \st{\rm{Cat}(D)}_N}{||a \st{\rm{Cat}(D)}_N||} = \frac{1}{\sqrt
2}\big{[} \sqrt{1+\frac{D}{N}} \st{\tilde N :\frac{D-1}{2}}_z \nonumber \\
&+ \sqrt{1-\frac{D}{N}} \st{\tilde N : -\frac{D+1}{2}}_z \big{]}, \\
\st{\tilde \Psi_B}_{\tilde N} =& \frac{b \st{\rm{Cat}(D)}_N}{||b
\st{\rm{Cat}(D)}_N||} = \frac{1}{\sqrt
2}\big{[}\sqrt{1-\frac{D}{N}} \st{\tilde N : \frac{D+1}{2}}_z \nonumber \\
&+ \sqrt{1+\frac{D}{N}} \st{\tilde N : -\frac{D-1}{2}}_z \big{]}.
\end{eqnarray}

Recently Savage {\it et al.} \cite{Savage4Mode} discussed the
effects of decoherence due to the presence of non-condensed atoms.
For low temperatures the primary effect is to introduce
phase-damping, which was taken into account by a four mode model
in \cite{Savage4Mode}. They showed that decoherence effects are
negligible for $N \approx 200$, which indicates that the present
scheme could be a robust implementation for creating MES.

\subsection{Measurement}
\label{sec:measurement}

By measuring $S_z$ one obtains $P(n,t)$ as shown in
Fig.~\ref{fig_bar_pics}. This, however, does not guarantee the
state to be macroscopically entangled, since a completely
incoherent mixture
\begin{equation}
\rho=\frac{1}{2}\st{D/2}_z\stdag{D/2}+\frac{1}{2}\st{-D/2}_z\stdag{-D/2}
\end{equation}
gives the same result for $P(n,t)$. To distinguish between the two
cases one has either to do density matrix tomography (endoscopy)
\cite{ZollerTomography,Endoscopy} or measure the purity of the
system.

The latter can be realized by time reversal which can
experimentally be implemented by using the external field to apply
a series of pulses as shown in Appendix~\ref{AppendixA}. Replacing
$H$ by $-H$ after time $t_c$ and measuring $S_x$ after $2 t_c$ the
two cases can be distinguished. For the pure case one obtains
always the same outcome $+N/2$, while for a mixture multiple
outcomes are possible. Also the partial revival with $R>1/2$ could
be used to distinguish between the two cases without the difficulty
of implementing $-H$.

\subsection{Imperfections in the external field}
\label{sec:imperfections}

Experimentally it is possible to adjust the phase $\beta$ of the
$\pi/2$ pulses very precisely, while it is much more difficult to
exactly fulfill the condition of having a pulse are equal to
$\pi/2$. Therefore we investigate the influence of errors in the
pulse-area of the initial $\pi/2$-pulse only.

The radius of the Bloch-sphere is $N/2$, while the width of the
initial distribution is $\sqrt N / 2$. Thus we expect that the
required precision in the pulse area scales like $1 / \sqrt N$. We
confirmed this expectation by numerical simulation for small $N$
up to $N \approx 10^3$.

\section{Conclusion}
In summary, we have investigated schemes which allow the creation
of macroscopically entangled states with a distance $0<D<N$ on time
scales proportional to $1/N$. Within a semiclassical approximation
we obtained analytical estimates for $D$ and the time $t_c$, which
are in excellent agreement with the numerical results. We also
showed that the fidelity of these states can be improved
significantly by one axis pre-squeezing. We estimated the effect of
imperfections and decoherence, and believe that the presented
scheme might be used to produce macroscopically entangled states
with present state of the art technology.

\acknowledgments This work was supported by the Austrian Science
Foundation and the European Union projects under the contracts
No.~HPRN-CT-2000-00121 and No.~HPRN-CT-2000-00125.

\appendix
\section{Implementing the Hamiltonian $-H$}
\label{AppendixA}

First we investigate the time-evolution of $H$ given in
Eq.~\eqref{HamiltonianAngularMomentum} for $\Omega \gg \chi N$.
Then we show how a continuous sequence of pulses of the external
field can be used to generate $-H$ which is needed to implement the
measurement process discussed in Sec.~\ref{sec:measurement}. For
simplicity we consider the case $-H(\Omega=0)=-\chi S_z^2$.

For $|\Omega| \gg |\chi| N$, one can neglect $\chi S_z^{2}$ in
(\ref{HamiltonianAngularMomentum}). A pulse of angle $\gamma$ and
phase $\beta$ is characterized by
\begin{equation}
\int_{-\infty}^{+\infty}|\Omega(t)| dt = \gamma > 0 \qquad
\Omega(t) = |\Omega(t)| e^{i \beta} \nonumber
\end{equation}
and implements the following time-evolution
\begin{equation}
U_\beta(\gamma)=e^{-i \gamma [\cos(\beta)S_x-\sin(\beta)S_y]}.
\end{equation}
Its effect corresponds geometrically to a counterclockwise
rotation around the unit-vector
 $\mathbf{n}=(\cos(\beta),-\sin(\beta),0)$ about an angle $\gamma$.
The inverse is found by $U_\beta(\gamma)^\dag = U_{\beta}(-\gamma)
= U_{\beta+\pi}(\gamma)$. Since a rotation about $\gamma$ around
the z-axis corresponds to three pulses
\begin{equation}
\begin{split}
R_z(\gamma)=&e^{-i\gamma S_z}
=U_0\left(\frac{\pi}{2}\right)U_{-\frac{\pi}{2}}(\gamma)U_{\pi}\left(\frac{\pi}{2}\right),
   \\ R_z ( \gamma ) ^ {\dag} =& R_z ( - \gamma ) = U_0 \left( \frac{\pi}{2} \right) U_{\frac{\pi}{2}} (\gamma) U_{\pi}
   \left(\frac{\pi}{2}\right),
\end{split}
\end{equation}
a rotation around an arbitrary vector
$\mathbf{n}=(\sin\theta\cos\phi,\sin\theta\sin\phi,\cos\theta)$
about an angle $\gamma$ is given by
\begin{equation}
\begin{split}
R_\mathbf{n}(\gamma)= e^{-i \gamma \mathbf{n}\cdot\mathbf{S}} =
U_{-\phi-\frac{\pi}{2}}\left(\frac{\theta}{2}\right)R_z(-\phi)U_{-\phi+\frac{\pi}{2}}\left(\frac{\theta}{2}\right).
\end{split}
\end{equation}

The time evolution operator $U$ at time $t=M \tau$ after $M$
arbitrary rotations is given by
\begin{eqnarray}
U(t)=U_{\rm{ps}}(\tau)^M=\prod_{k=1}^{M}U_{\rm{ps}}(\tau) \quad \mbox{with} \nonumber \\
U_{\rm{ps}}(\tau)=\prod_{j=1}^{J} R_{\mathbf{n}_j}(\gamma_j)^\dag
e^{-i \chi S_z^2 \tau/J} R_{\mathbf{n}_j}(\gamma_j) =e^{-i
H_{\rm{eff}} \tau}. \nonumber \label{MultiPulseSeq}
\end{eqnarray}

By combining two rotations about around axes ${\bf n}_1,{\bf n}_2$
($J=2$) we obtain $H_{\rm eff}=\chi (S_{{\bf n}_1}^2 +S_{{\bf
n}_2}^2)/2$, since
\begin{eqnarray}
U_{\rm{ps}}(\tau) =& e^{-i \chi S_{{\bf n}_1}^2 \tau/2}e^{-i \chi S_{{\bf n}_2}^2 \tau/2} \approx \left(1-i\chi S_{{\bf n}_1}^2\tau/2\right)\times \nonumber\\
& \left(1-i\chi S_{{\bf n}_1}^2\tau/2\right) \approx e^{-i\chi
\left( S_{{\bf n}_1}^2 + S_{{\bf n}_2}^2\right) \tau/2}
\end{eqnarray}
to first order in $\tau$. Thus $H_{\rm{eff}}=-H$ can be implemented
(up to a constant) by choosing ${\bf n}_1=(1,0,0)$,${\bf
n}_2=(0,1,0)$ and making use of the identity
\begin{equation}
\mathbf{S}^2=S_x^{2}+S_y^{2}+S_z^{2}=\frac N 2 \frac{N+2} 2.
\label{SpinIdentity}
\end{equation}

\section{Semiclassical Calculations}
\label{AppendixB}

In this appendix we will derive the approximations of
Sec.~\ref{sec:SCRes}. First we derive the maximal distance $D$
given in Eq.~\eqref{eq:distance} for $\sigma \rightarrow 0$, then
we calculate the time $t_c$ given in Eq.~\eqref{eq:timescale}
needed to create the macroscopic superposition states. For
 simplicity we restrict $\theta,\phi$
to the intervals $0 \leq \theta \leq \pi$, $-\pi \leq \phi \leq
\pi$ in Eqs.~(\ref{eq:separatrix},\ref{classicalVF}). Therefore we
have to distinguish for the $\pm$ sign in the second relation of
Eq.~\eqref{eq:separatrix} for the two regimes of interest, i.e.~for
$\omega \leq 1$ only the $+$ sign holds, whereas for $\omega > 1$
both signs hold.

For $\sigma \rightarrow 0$ $D$ corresponds to the maximum
separation along the separatrix for given $\phi$, i.e.
\begin{equation}
D= \begin{array}{c}\\{\rm max}\\\phi[\theta]\end{array} N \cos
\theta.
\end{equation}
Using Eq.~\eqref{eq:separatrix} we obtain the maximum value for
$\omega \leq 1$ at $\phi=\pi$ as
\begin{eqnarray}
D =& N \cos \theta^{+}[\phi=\pi] = N \sqrt{1-(\sin \theta^{+}[\phi=\pi])^2}\nonumber\\
=& N \sqrt{1- \left( \frac{-\omega + |\omega-2|}{2} \right)^2}
\nonumber = N \sqrt{\omega (2-\omega)}.
\end{eqnarray}
Similarly for $1 \leq \omega \leq 2$ we obtain  three values for
$\theta$ at $\phi=0$ given by $\theta_0=0$ and the points of
maximum separation $\theta_{\pm}^{-}$ for the minus sign in
Eq.~\eqref{eq:separatrix}. The separation $\gamma$ is given by
\begin{eqnarray}
\gamma =& \theta_{+}^{-}-\theta_{-}^{-} = \pi -2 \arcsin
\left(-\frac{\omega}{2}-\frac{|\omega-2|}{2}\right)\nonumber\\ =&
\pi - 2 \arcsin(\omega-1) = 2 \arcsin \left(\sqrt{\omega
(2-\omega)}\right).
\end{eqnarray}

Next we calculate the time $t_c$ given in Eq.~\eqref{eq:timescale}.
Using Eq.~\eqref{classicalVF} and the conservation of energy as
stated in Sec.~\ref{sec:Distance} we obtain $d\theta/dt$ along the
separatrix for $0\leq\theta\leq\pi/2,0\leq\phi\leq\pi$, as
\begin{equation}
\frac{d\theta}{dt}=-\chi \frac N 2 \sin \phi = -\chi \frac N 2
\sqrt{1-\left(\frac{\omega - \cos^2 \theta}{\omega \sin
\theta}\right)^2}.
\end{equation}
Integrating this ordinary differential equation gives
\begin{eqnarray}
\chi t = - \frac{2}{N \omega} \int_{\theta(0)}^{\theta(t)} d\theta'
\left[ 1- \left(\frac{1-\omega^{-1}\cos^2\theta}{\sin
\theta}\right)^2 \right]^{-1/2} \nonumber \\ = \frac{2}{N \omega}
\int_{z(0)}^{z(t)}
dz \left[ 1-z^2 - (1-\omega^{-1} z^2)^2 \right]^{-1/2} \nonumber \\
= \frac{2}{N \sqrt{\omega (2-\omega)}} \log \left[
\frac{z(t)}{z(0)}\frac{1 + \sqrt{1-z(0)^2/\omega(2 -\omega)}}{1 +
\sqrt{1-z(t)^2/\omega(2 -\omega)}} \right],\label{integral}
\end{eqnarray}
where $z(t)\equiv\cos \theta(t)$. For a point on the separatrix at
a distance $\sigma$ from $\theta=\pi/2,\phi=0$, i.e. the peak of
the initial distribution, $z(0)$ is obtained from the relation
\begin{eqnarray}
\sigma^2 &=& 4\left({\bf S}-{\bf S}(0)\right)^2/N^2 = 2 (1 - \sin
\theta \cos \phi )\nonumber\\ &=& 2 \cos^2 \theta / \omega = 2
z(0)^2/\omega. \label{initial}
\end{eqnarray}
The time needed to travel from $z(0)=\sqrt{\omega/2\sigma}$ to
$z(t)=D/N=\sqrt{\omega (2-\omega)}$, i.e.~the point of maximum
separation of the two number state [gaussian distributions] forming
the macroscopic superposition state
(\ref{CatState1},\ref{CatState2}) is then obtained using
Eq.~\eqref{integral} as
\begin{eqnarray}
\chi t &=& \frac{2}{N \sqrt{\omega (2-\omega)}}\log \frac{1 +
\sqrt{1-\sigma^2/2 (2-\omega)}}{\sqrt{\sigma^2/2(2-\omega)}} \nonumber \\
&=& \frac{2 \log \left[ \sqrt{2(2-\omega)/\sigma^2} +
\sqrt{2(2-\omega)/\sigma^2-1}\right]}{N \sqrt{\omega (2-\omega)}}.
\end{eqnarray}
Finally using $\sigma=1/\sqrt{N}$ we obtain
Eq.~\eqref{eq:timescale}.


\end{document}